\documentstyle[preprint,aps,prl,epsf]{revtex}

\begin{document}
\draft
\title{The hyperfine structure of highly charged
$^{238}_{\ 92}$U ions with rotationally excited nuclei}
\author{L. N. Labzowsky$^{1,2,3}$, A. V. Nefiodov$^{1,2,3}$, 
G. Plunien$^4$, G. Soff$^4$, and D. Liesen$^5$}
\address{$^1$Max-Plank-Institut f\"ur Physik komplexer Systeme,
N\"othnitzer Strasse 38, D-01187 Dresden, Germany \\
$^2$St.~Petersburg State University, 198904 St.~Petersburg, 
Russia \\
$^3$Petersburg Nuclear Physics Institute, 188350 Gatchina, 
St.~Petersburg, Russia \\
$^4$Institut f\"ur Theoretische Physik, Technische  Universit\"at
Dresden, Mommsenstrasse 13, D-01062  Dresden, Germany \\
$^5$Gesellschaft f\"ur Schwerionenforschung (GSI), Planckstrasse 1,
D-64291 Darmstadt, Germany}
\date{Received \today}
\maketitle

\widetext
\begin{abstract}
The hyperfine structure (hfs) of electron levels of $^{238}_{\ 92}$U
ions with the nucleus excited in the low-lying rotational 
$2^+$ state with an energy $E_{2^+} = 44.91$ keV is investigated. 
In hydrogenlike uranium, the hfs splitting for the $1s_{1/2}$-ground  
state of the electron constitutes $1.8$ eV. The hyperfine-quenched (hfq) 
lifetime of the $1s2p$~${}^3P_0$ state has been calculated for heliumlike
$^{238}_{\ 92}$U and was found to be two orders of magnitude smaller
than for the ion with the nucleus in the ground state. The possibility 
of a precise determination of the nuclear $g_r$ factor for the rotational 
$2^+$ state by measurements of the hfq lifetime is discussed.
\end{abstract}

\narrowtext
Atomic hfs experiments and pure nuclear measurements (pionic
scattering etc.) are two supplementary ways for obtaining 
informations on nuclear moments. However, up to now the atomic hfs
experiments have been performed exclusively for atoms or
ions with nuclei in the ground state. In this paper we point out 
that experiments are also feasible for highly charged ions 
with rotationally excited even-$A$ nuclei.

The low-lying excited rotational state of the $^{238}_{\ 92}$U nucleus
with excitation energy $E_{2^+} = 44.91$ keV plays an important
role in recent accurate calculations of the Lamb shift for
highly charged uranium ions. In particular, for these ions the
contribution of the $2^+$ state dominates in calculations of
nuclear polarization shifts \cite{1,3,4,5}. However, in such calculations 
this state enters as a virtual nuclear excitation. In the present 
work we consider the situation of a real excitation of the $2^+$ state 
due to the interaction of the incoming atomic uranium beam with a target 
in beam-foil experiments. 

The empirical energy spectrum of a rotationally excited nucleus in the
ground-state band fits well to the formula 
($\hbar=c=1$) \cite{6,7}
\begin{equation}                                             
E_{I} = {\cal A} I(I+1).
\label{eq1}                                         
\end{equation}
In Eq.~(\ref{eq1}), $I$ denotes the total angular momentum of the rotating 
nucleus and ${\cal A}$ is the rotational constant. The latter can be related 
to the moment of inertia ${\cal I}$ of the nucleus according to 
${\cal A} =  1/2{\cal I}$.
The lowest rotational excitation in $^{238}_{\ 92}$U is the electric 
quadrupole transition $0^+ \rightarrow
2^+$ within the ground-state band with an excitation energy
$E_{2^+} = 44.91$ keV. Fitting this energy to Eq.~(\ref{eq1}) yields
the rotational constant ${\cal A} \simeq 7.5$ keV \cite{8}.

A rotationally excited nucleus should have a magnetic moment 
$\bbox{\mu} = \mu'\bbox{n} + \mu_N g_r(\bbox{I}-\Omega\bbox{n})$
associated with its total angular momentum $\bbox{I}$  \cite{6,7}.
Here $\mu'\bbox{n}$ denotes the magnetic moment of the nonrotating
nucleus, $\bbox{n}$ represents the unit vector directed along the nuclear
axis, and $\mu_N$ is the nuclear magneton.
The ratio $g_r ={\cal I}_p/{\cal I}$ defines the gyromagnetic factor for the
rotation of the nucleus with ${\cal I}_p$ being the protonic part of
the total moment of inertia ${\cal I}$. The projection $\Omega= 
(\bbox{I}\bbox{n})$ characterizes the various rotational bands. 
After averaging over rotations the magnetic moment is directed along 
the conserving vector $\bbox{I}$:
\begin{equation}
\bbox{\hat{\mu}} = \frac{\mu}{I}\bbox{\hat{I}}
= \mu' \overline{\bbox{n}} +  \mu_N g_r
(\bbox{\hat{I}} - \Omega\overline{\bbox{n}}).
\label{eq3}                                            
\end{equation}
Multiplying Eq.~(\ref{eq3}) by $\bbox{\hat{I}}$ and passing over
to eigenvalues, we obtain
\begin{equation}
\mu = \mu'\frac{\Omega}{I+1} + \mu_N g_r \left( I -
\frac{\Omega^2}{I+1}\right).
\label{eq4}                                           
\end{equation}
Thus in highly charged ions with $\mu'=0$ but rotationally excited 
nuclei a hyperfine structure splitting of levels should arise. 

In the point-nucleus approximation  the hfs magnetic-dipole interaction 
operator $\hat H_{\rm hfs}$ can be written in the form:
\begin{equation}
\hat H_{\rm hfs}(\bbox{r})= e\bbox{\hat{\mu}}\frac{\left[
\bbox{\alpha} \times \bbox{r}\right]}{r^3},
\label{eq5}                                          
\end{equation}
where the nuclear magnetic moment is defined by Eq.~(\ref{eq3}),
$e$ is the electron charge $(e<0)$, and $\bbox{\alpha}$ and
$\bbox{r}$ are the Dirac matrices and the spatial coordinate for the
atomic electron, respectively. For a spinless nucleus in the 
ground-state band $(\Omega=0)$ the expression (\ref{eq4}) yields
$\mu =\mu_N\,g_r I$.

The hfs correction to the energy levels of hydrogen-like
$^{238}_{\ 92}$U ion is defined by the standard Land\'{e}
expression $\Delta E_{\rm hfs}(F) =Ca/2$, 
where the cosine factor is $C = F(F+1)-I(I+1)-j(j+1)$, $j$ is the
total electron angular momentum, $F$ is the total angular
momentum of an ion, and the hfs constant $a$ is determined by
\begin{equation}
a= g_r\frac{\alpha}{m_p}\frac{\kappa}{j(j+1)}\int_0^\infty
\frac{dr} {r^2} \, P_{nlj}(r)Q_{nlj}(r).
\label{eq7}                                           
\end{equation}
Here $\alpha = e^2$ is the fine-structure constant, $m_p$ is 
the proton mass,  
and $P_{nlj}(r)$ and $Q_{nlj}(r)$ are the upper and lower
radial components of the electron wave function characterized by the
principal quantum number $n$ and the relativistic quantum number
$\kappa=(l-j)(2j+1)$. Employing analytical results from Refs.~\cite{9,10},
one finds
\begin{equation}
a=\alpha(\alpha Z)^3 g_r\frac{m_e^2}{m_p}\frac{\kappa \left[
2\kappa(\gamma + n_r) - N\right]}{j(j+1)N^4\gamma(4\gamma^2
-1)}(1 - \delta_{nlj}),
\label{eq8}                                           
\end{equation}
where $\gamma = \sqrt{\kappa^2 -(\alpha Z)^2}$, $N =
\sqrt{n_r(2\gamma + n_r) +\kappa^2}$, $m_e$ is the electron
mass, $Z$ is the number of protons, and $n_r$ is the radial
quantum number ($n=n_r +|\kappa|$). The
correction $\delta_{nlj}$ accounts for the finite nuclear charge
distribution. The nuclear magnetization distribution correction
as well as QED corrections are rather small and therefore
negligible. For electron states with $j=1/2$, the hfs splitting 
$\Delta E_{\mu}$ between the states with $F=I+1/2$ and $F=I-1/2$ 
is just $\Delta E_{\mu} =(I+1/2)a$.
Assuming homogeneous nuclear charge and mass
distributions one obtaines $g_r=Z/A$ for a nucleus with 
mass number $A$. Thus from atomic hfs experiments with 
rotationally excited nuclei one can deduce directly the
deviation of the empirical $g_r$ factor from the $Z/A$ approximation. 
To our knowledge, 
$g_r$ factors for $^{238}$U have been determined only for 
rotational states with spin $I=6$ and higher by means of  
measurements of the precession angles in transient magnetic fields by 
$\gamma$-ray - particle coincidences \cite{10a}. 
These measurements cannot be performed in the case of the highly 
converted, low-lying $2^+$ and $4^+$ nuclear states. 

\begin{figure}[h]
\centerline{\mbox{\epsfxsize=8cm \epsffile{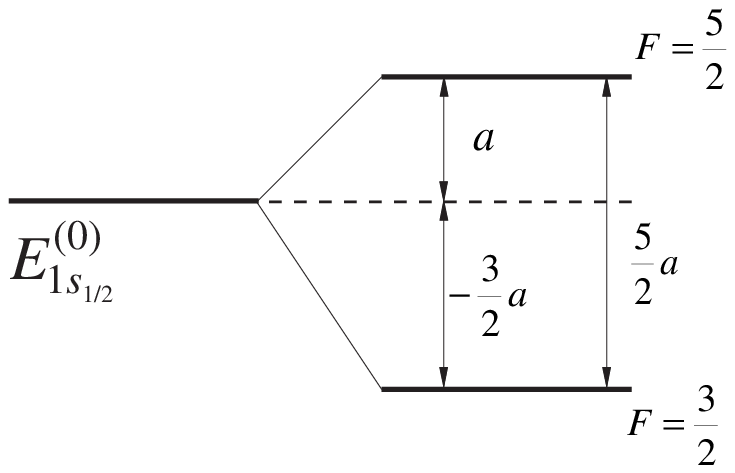}}}
\vspace*{1cm}
\caption{The hyperfine splitting $\Delta E_{\mu}$ of the ground state
of hydrogenlike $^{238}_{\ 92}$U with the nucleus in the rotationally
excited $2^+$ state.}
\label{fig1}
\end{figure}

For the $1s_{1/2}$-ground state of hydrogenlike 
$^{238}_{\ 92}$U the splitting is indicated in Fig.~\ref{fig1}.
The value of $E^{(0)}_{1s_{1/2}}$ corresponds to the ground-state
energy level of the uranium ion with the unexcited
nucleus. The uncertainty of the value $E^{(0)}_{1s_{1/2}}$ is
determined by the Lamb shift calculations. The present
theoretical value for the Lamb shift is $\Delta E^{\rm
(th)}_{1s_{1/2}} = 464.7 \pm 1.0$ eV \cite{11}, while the
experimental value is $\Delta E^{\rm (exp)}_{1s_{1/2}} = 470
\pm 16$ eV \cite{12}. The evaluation of the hfs constant
(\ref{eq8}) for the ground state of $^{238}_{\ 92}$U$^{91+}$
with $g_r=Z/A$ yields $a=0.89$ eV for a point nucleus and
$a=0.72$ eV for an extended one. In the latter case, the finite-size 
correction $\delta_{1s}=0.19$ has been approximated according to
results obtained in Ref.~\cite{10}. Then one finds a hyperfine
splitting $\Delta E_{\mu} = 1.8$ eV ($\Delta \lambda = 0.69$
$\mu$m), which is well resolvable within the accuracy of 1 eV 
envisaged for the near-future level shift measurements in hydrogenlike
uranium \cite{12}.

The lifetime $\tau_{2^+}$ of the excited rotational state $2^+$
of the $^{238}_{\ 92}$U nucleus can be obtained from the known empirical 
value for the reduced transition probability $B(E2; 0^+
\rightarrow 2^+)= 12.3$ $e^2 b^2$ \cite{8}, where {\it b} denotes
barn. This leads to $\tau_{2^+} \simeq 10^2$ ns. One should point out 
that the much smaller value $\tau_{2^+} \simeq 10^2$ ps given in the
literature \cite{8} corresponds to neutral uranium atoms and
is due to the internal conversion process. The latter decay
channel is absent in H- and He-like $^{238}_{\ 92}$U ions (see
also \cite{13,14}). The time $\tau_{2^+}$ is large enough to
consider the magnetic interaction between the electron in
hydrogenlike uranium and the rotating nucleus as a stationary
problem. The time of revolution $\tau_{{\rm rot}}$ associated
with this nuclear excitation can be deduced from
\[ E_{2^+} = \frac{1}{2}{\cal I}\omega_{\rm rot}^2 = \frac{1}
{4{\cal A}}\left(\frac{2\pi}{\tau_{\rm rot}}\right)^2
\]
yielding $\tau_{\rm rot}\simeq 10^{-4}$ fs, which is negligibly small
compared to $\tau_{2^+}$. 

Let us now discuss the possibility of an
experimental verification of this effect using the SIS/ESR
facility at GSI in Darmstadt. We consider a beam of bare uranium ions
with typical kinetic energy  $E_{\rm kin} \simeq 320$ MeV/u 
which corresponds to a velocity $v \simeq 0.67 c$. 
From the lifetime $\tau_{2^+} \simeq 10^2$ ns it follows that the
decay length of the $2^+$ state is larger than $25$~m behind the foil.
The number of ions $n_i$ with
rotationally excited nuclei that can be prepared per second is given by 
$n_i = J \sigma {\cal N}$, where $J$ denotes the intensity of the ion beam,
$\sigma$ is the cross section for excitating 
the nucleus inside the foil, and ${\cal N}$ is the number of foil
atoms per unit area.

The Coulomb excitation cross section $\sigma$
for uranium nuclei in collisions with nuclei of the carbon foil
can be estimated within the framework of the equivalent photon
method as described in Ref.~\cite{15}. Assuming that only the rotational 
$2^+$ state of $^{238}$U with the energy $E_{2^+} = 44.91$ keV is excited, 
the photonuclear absorption cross section may be approximated by
$\sigma^{E2}_{\gamma}(\varepsilon) \simeq \frac{4\pi^3}{75}
\left(\frac{\varepsilon}{\hbar c}\right)^3 B(E2)\,\delta(\varepsilon
-E_{2^+})$. This approximation is legitemized by the huge value of 
the reduced transition strength $B(E2)$ of the rotational $2^+$ 
state as the most dominant collective nuclear excitation of the 
$^{238}$U isotope. Then the total cross section $\sigma$ results as 
$\sigma \simeq \int \frac{{\rm d}\varepsilon}{\varepsilon} \,
n^{E2}(\varepsilon)\sigma^{E2}_{\gamma}(\varepsilon)$, where
$n^{E2}(\varepsilon)$ denotes the number of equivalent photons.
The adiabaticity parameter involved in the problem is equal to 
$\xi=\frac{b_{\rm min}E_{2^+}}{\hbar c\beta
\gamma}$, where $\beta=v/c \simeq 0.67$, $\gamma =
(1-\beta^2)^{-1/2}$, and $b_{\rm min}$ is the minimum impact
parameter taken as the sum of the two nuclear radii. Since $\xi
\simeq 2.6\times 10^{-3}$ is quite small, the number of
equivalent photons can be approximated by 
$n^{E2}(E_{2^+})\simeq \frac{4\alpha Z_f^2}{\pi \gamma^2\xi^2
\beta^4}$, where $Z_f$ is the nuclear charge number of the foil atoms. 
Finally, we obtain $\sigma \simeq 10.7$ fm${}^2$.
Taking the value ${\cal N} \simeq 0.5\times 10^{20}$ ${\rm cm}^{-2}$
for a typical carbon foil density $\rho=1 $ mg/${\rm cm}^2$ together with a  
characteristic intensity $J\simeq 10^{10}$ ions/s, one
finds $n_i \simeq 0.5 \times 10^{5}$ ions/s. 
Only a fraction of about  $0.5\times 10^{-5}$ 
of the primary ions are ions with excited nuclei in the 
rotational $2^+$ state. However, even if all electrons captured in the foil 
are supposed to decay to the $1s_{1/2}$-ground state by the emission 
of Lyman radiation, a direct measurement of the hfs splitting is at present 
prohibited by the low efficiency ($\sim 10^{-8}$) of the required
high-resolution  spectrometers. 

\begin{figure}[h]
\centerline{\mbox{\epsfxsize=12cm \epsffile{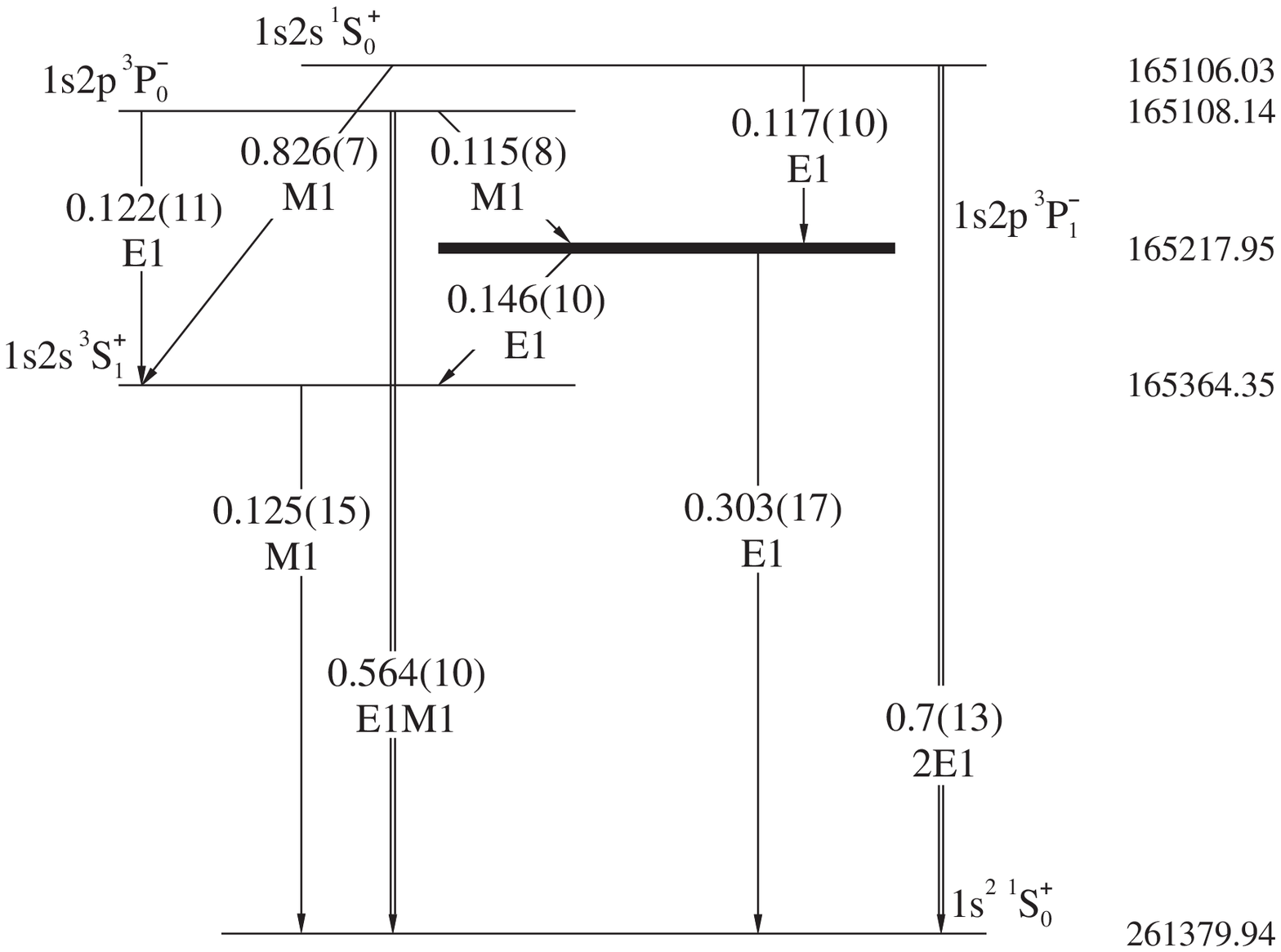}}}
\vspace*{1cm}
\caption{Energy level scheme of the first excited states of
heliumlike
$^{238}_{\ 92}$U with the nucleus in the ground
state. The partial probabilities of radiative
transitions are given in s$^{-1}$. Numbers in brackets indicate
powers of 10. 
The large radiative width for the $1s2p$~$^3P_1$ state is
indicated as a bold line. 
The double lines denote two-photon transitions. 
Numbers on the right-hand side indicate the binding
energies in eV.}
\label{fig2}
\end{figure}

Still there exists another possibility for the observation
of the effect. It is based on the measurement of the
hfq lifetime of the metastable  $1s2p$~${}^3P_0$
level in  $^{238}_{\ 92}$U$^{90+}$  ions. 
This effect was observed in Refs.~\cite{16,16a,16b,16c,17,18} for the
isoelectronic sequence of heliumlike ions with non-zero
nuclear spin. The level scheme of the first excited states in the 
$^{238}_{\ 92}$U$^{90+}$ ion without taking into accout the hfq 
decay channels is depicted in
Fig.~\ref{fig2}. The energy values and partial transition
probabilities were calculated within the framework of the
multiconfigurational Dirac-Fock method (MCDF) \cite{19}. The energies
of the electron levels include the radiative \cite{20,21}
(electron self-energy and vacuum polarization) and the exact
one-photon exchange corrections. The E1M1 two-photon transition 
rate has been calculated by Drake \cite{25} and the 2E1 decay rate is 
taken from Ref.~\cite{26}.

Hyperfine quenching of the metastable $2^3P_0$ state results from a 
mixing with the short-lived $2^3P_1$ state by the hyperfine
interaction (\ref{eq5}). The partial widths $\Gamma_J$ ($J=0,1$) for
$2^3P_J$ levels due to the radiative E1 transitions to the
ground $1^1S_0$ state are related by 
\begin{equation}
\Gamma_0 = \eta^2 \Gamma_1,
\label{eq9b} 
\end{equation}                                         
where the mixing coefficient $\eta$ is defined by
\begin{equation}
\eta= \sum_{i=1}^{2}\frac{\langle 2^3P_0|\hat H_{\rm hfs}
(\bbox{r}_i)| 2^3P_1\rangle} {E_{2^3P_0} -E_{2^3P_1}}
\label{eq10}                                           
\end{equation}
and the rotationally-induced hyperfine interaction operators
$\hat H_{\rm hfs}(\bbox{r}_i)$ are given by Eq.~(\ref{eq5}).
The coefficient $\eta$ can be expressed directly through the
$g_r$ factor. Performing the integrations over the angles, the matrix
element in expression (\ref{eq10}) reads 
\widetext
\begin{eqnarray}
\langle 2^3P_0|\sum_{i=1}^{2}\hat H_{\rm hfs}(\bbox{r}_i)|
2^3P_1\rangle &=& g_r\frac{2 \alpha}{3 m_p}\sqrt{I(I+1)}
\int_0^\infty \frac{dr} {r^2} \, \left[ P_{1s}(r)Q_{1s}(r) +
P_{2p_{1/2}}(r)Q_{2p_{1/2}}(r) \right] \nonumber\\
&=& g_r\alpha (\alpha Z)^3  \sqrt{I(I+1)}\frac{m_e^2}{m_p}
\left[ \frac{(2\gamma +2 - \sqrt{2\gamma +2})}{(2\gamma +2)^2 \gamma
(4\gamma^2 -1)}(1-\delta_{2p}) -
\frac{(1 - \delta_{1s})}{\gamma (2\gamma -1)}\right], \nonumber
\end{eqnarray}
\narrowtext
\noindent 
where $\gamma =\sqrt{1- (\alpha Z)^2}$. 

For the $^{238}_{\ 92}$U$^{90+}$ ion, the mixing coefficient
(\ref{eq10}) has been calculated for $g_r=Z/A$ in the framework of 
the MCDF approach to yield 
\[
\eta = - \frac{0.764 \  \mbox{\rm eV}}
{E_{2^3P_0} -E_{2^3P_1}} = 0.696\times 10^{-2}.
\]
It leads to the appearance of an additional contribution to
the radiative width of the $2^3P_0$ level, that turns out to be
$0.147 \times 10^{13}$ s$^{-1}$. As a result, the lifetime of
the $2^3P_0$ level is diminished from 56 ps to 0.67 ps, which
corresponds to a decay length of about $0.18$ mm in the laboratory.
We should emphasize that there will be no background from the ions 
with unexcited
nuclei, since in those ions the one-photon transition $2^3P_0
\rightarrow 1^1S_0$ with the energy $\omega_0 \simeq 96.271$ keV
is absolutely forbidden. The lifetime of 56 ps for 
ions in the $2^3P_0$ state arises from  the transition $2^3P_0
\rightarrow 2^3S_1$ with the energy $\omega_1 \simeq 256.21$ eV,
which is far away from $\omega_0$ ($70\%$ of the total width) as well 
as from the two-photon transition $2^3P_0 \rightarrow 1^1S_0$
($30\%$ of the total width). Both of these transitions cannot give
any background contribution in the proposed experiment. The transition
$2^3P_1 \rightarrow 1^1S_0$ with the energy of about $96.162$ keV
does also not contribute to the background, since the level $2^3P_1$ has
a lifetime of 0.033 fs and hence decays already inside of the foil.
In order to obtain clean spectra without loss of efficiency, a 
measurement of coincidences between heliumlike ions and photons
should be performed. Thus, the observation of the predicted effect 
becomes feasible utilizing the beam-foil time-of-flight technique.
In view of Eq.~(\ref{eq9b}), the accuracy of the determination of the $g_r$ 
factor is even two times better than the accuracy of the measured hfq 
lifetime, which in this region can be expected to be at the level of 
about $1$\%.

\acknowledgments
The authors are indebted to I.M. Band, V.I. Isakov, and K. H. Speidel for
helpful discussions. L. L. and A. N. are grateful to the Technische 
Universit\"at Dresden and the Max-Planck-Institut f\"ur Physik 
komplexer Systeme (MPI) for the hospitality and for financial 
support from the MPI, DFG, and the RFFI
(grant no. 99-02-18526). G. S. and
G. P. acknowledge financial support from BMBF, DAAD, DFG, and
GSI.

\newpage

\newpage

\end{document}